\documentstyle[twoside,fleqn,espcrc2,epsf]{article}

\newcommand{\AmS}{{\protect\the\textfont2
  A\kern-.1667em\lower.5ex\hbox{M}\kern-.125emS}}

\title{Results from the AMANDA High Energy Neutrino Detector}

\author{
{\footnotesize
\noindent
E. Andres$^{8}$, 
P. Askebjer$^{4}$, 
X. Bai$^{1}$, 
G. Barouch$^{8}$, 
S. W. Barwick$^{6}$, 
R.C. Bay$^{5}$, 
K.-H. Becker$^{13}$, 
L. Bergstr\"om$^{4}$, 
D. Bertrand$^{10}$, 
A. Biron$^{2}$
% ...
\newline
\newline
\noindent \textit{
(1)Bartol Research Institute, University of Delaware, Newark, DE 19716, USA\\
(2) DESY-Zeuthen,D-15735, Zeuthen, Germany\\
(3) Dept. of Technology, University of Kalmar, S-39129, Kalmar, Sweden\\
(4) Dept. of Physics, Stockholm University, S-11385, Stockholm, Sweden\\
(5) Dept. of Physics, University of California, Berkeley, CA 94720 USA\\
(6) Dept. of Physics and Astronomy, University of California, Irvine, CA 92697 USA\\
   (7) Dept. of Physics and Astronomy, University of Pennsylvania, Philadelphia, PA 19104 USA\\
   (8) Dept. of Physics, University of Wisconsin, Madison,  WI 53706 USA\\
   (9) Lawrence Berkeley National Laboratory, Berkeley, CA 94720 USA\\
   (10) Brussels Free University, B-1050, Brussels, Belgium\\
   (11) Dept. of Radiation Sciences, University of Uppsala, S-75121, Uppsala, Sweden\\
   (12) Institute of Physics, University of Mainz, D-55099, Mainz Germany\\
   (13) Fachbereich 8 Physik, BUGH Wuppertal, D-42097 Wuppertal Germany
}
}
}

\begin{document}  

\begin{abstract}
This paper briefly summarizes the search for astronomical sources of high-energy neutrinos using the AMANDA-B10 detector.  The complete data set from 1997 was analyzed. For $E_{\mu}>10$ TeV, the detector exceeds 10,000 $m^2$ in effective area between declinations of 25 and 90 degrees. Neutrinos generated in the atmosphere by cosmic ray interactions were used to verify the overall sensitivity of the detector.  The absolute pointing accuracy and angular resolution has been confirmed by the analysis of coincident events between the SPASE air shower array and the AMANDA detector. Preliminary flux limits from point source candidates are presented.  For declinations larger than +45 degrees, our results compare favorably to existing limits for sources in the Southern sky. We also present the current status of the searches for high energy neutrino emission from diffusely distributed sources, GRBs, and WIMPs from the center of the earth. 
\vspace{1pc}
\end{abstract}

\maketitle

\section{INTRODUCTION}

The AMANDA-B10 high energy neutrino detector was constructed between 1500--2000 m
below the surface of the Antarctic ice sheet where the optical properties are suitable for track reconstruction\cite{ICRC99}. The instrumented volume forms a cylinder with outer diameter of 120 m. The surface electronics
are located within a kilometer of the Amundsen-Scott Research Station
at the geographic south pole. The detector was commissioned in
February 1997\cite{Aman99}\cite{TAUP99}, and initial scientific results were presented at the XXIVth International Cosmic Ray Conference\cite{ICRC99}.  Reconstruction methods and detector calibration techniques are introduced in reference \cite{Andres00}.

AMANDA-B10 consists of 302 optical modules (OMs) that contain
an 8 inch diameter photomultiplier tube controlled by passive electronics
and housed in a glass pressure vessel. They are connected to the surface by
an electrical cable that provides high voltage and transmits the signals from
the OM.  The simple, reliable system architecture is responsible for the
low fraction of OM failure ($<$ 10\% after several years of operation, although most of the failures occur within a week of deployment).

In January, 2000, AMANDA-II was completed.  It consists of 19 strings with at total of 677 OMs arranged in concentric circles, with the ten strings from AMANDA-B10 forming the central core of the new detector.  New surface electronics consolidates several triggering functions and adds functionality. New scalers were installed that provides millisecond resolution - important for Supernova studies. Several technologies were deployed to evaluate their utility and readiness for future expansion to larger systems.

\begin{figure} [ht!]

\epsfxsize=8cm\epsffile{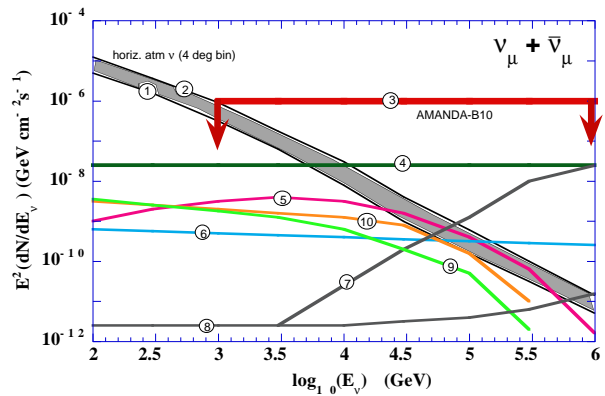}
\caption{ Survey of $\nu+\overline{\nu}$ flux predictions from cosmic accelerators of high energy neutrinos (adopted from the recent review by Learned and Mannheim\cite{ML00}. The atmospheric
neutrino fluxes are from ref. \cite{atm_nu_pred}, for both vertical (1) and horizontal (2) fluxes within a circle defined by a half angle of 4 degrees ( labeled "4 deg bin"). The curves do not include the normalization uncertainty, possibly 20\% in magnitude. Numbered lines: (3) AMANDA-B10 limit reported here; (4) Ref. \cite{Nellen} for the core emission from 3C273 due to pp interactions.  It also represents neutrino emission from the AGN Blazar Mk501 during 1997 if it emits half of its TeV gamma ray flux in neutrinos. (5) Crab Nebula. Model I from ref. \cite{Bednarek97}, (6) Coma cluster according to ref. \cite{Cola}, (7) core emission from 3C273 due to p-$\gamma$ interactions\cite{SS96}, (8)model\cite{Mannheim93} for the relativistic jet of 3C273 including p-p and p-$\gamma$ interactions, and Supernova remnant gamma-Cygni (9) and IC 444 (10) according to ref.\cite{Gaisser98}.  The energy bounds on the AMANDA limit are restricted to the approximate region of sensitivity of the detector.}
\label{fig:theory}
\end{figure}

The search for astrophysical sources of high energy ($E_{\nu}>$ 1 TeV)
 neutrinos is one of the central missions of the AMANDA detector. Figure~\ref{fig:theory} provides a survey of model predictions for the flux of high energy neutrinos from point objects. It also contains the flux limit reported in this manuscript for sky bins with declinations greater than 30 degrees. Most theoretical models of potential astrophysical sources of neutrinos predict that the energy spectrum is very hard, approximately $E^{-2}$\cite{point_pred}. Due to the hard energy spectrum, the most probable energy of the detected neutrino is well above 1 TeV (typically 10-30 TeV). 

 In this report, we describe the status of a general search for continuous 
 emission from point sources in the northern sky, restricted to declination greater than +$5^{o}$. The observation of atmospheric neutrinos and detailed studies of downgoing atmospheric muons have been used to study the sensitivity of the detector.  The observed differences between observation and prediction are within $\pm$30\%, which is within the systematic uncertainty. By restricting the atmospheric neutrino analysis to a search cone with a half angle of $15^o$ about zenith, a search was made for neutrino emission from WIMP annihilation near the center of earth. Finally, we report preliminary results for a search for high energy neutrinos from Gamma Ray Bursts (GRBs).  Neutrino signal from GRBs should be correlated in direction and time, greatly simplifying background rejection.  All results reported here are based on data collected between April and October of 1997.  Due to limitations in the data acquisition and archiving system at that early phase of operation, the livetime ranges between 130 and 140 days, depending on the details of the analysis.

\section{ANALYSIS PROCEDURE - SEARCH FOR POINT SOURCES}
The various physics objectives are best addressed by specialized analyses, although most are closely related.  The search for sources of point emission can be used to illustrate the general methods. Trigger events are dominated by downgoing atmospheric muons, so analysis
techniques were developed to reject this background while retaining good
efficiency for upgoing neutrino-induced muons.  
Unlike many neutrino detectors, the effective sensitivity
depends strongly as a function of the background rejection requirements, which are considerably weaker for searches for point sources than diffuse emission. 

The analysis procedure utilizes two essential characteristics of the signal to simplify the analysis relative to atmospheric neutrino measurements. First, the sources are assumed to be point sources in the sky, so only events within a selected angular region are considered.  Secondly, we use the topological characteristics of spectrally hard neutrino signal to reject poorly reconstructed atmospheric muons and atmospheric neutrinos, both of which have softer spectra. Topological variables include an estimate of muon energy and an assessment of the spatial fluctuation of the detected signals in a given event.  The complete suite of variables was able to differentiate signal events from several classes of background topologies.

Monte Carlo based simulation programs determined the effective area for background and neutrino
-induced muons.  Several important results from these programs were tested by comparing the background simulation to the experimental data at various steps along the analysis chain. 

An iterative analysis procedure was developed to maximize the $S/\sqrt{BG}$, where the signal, S,  was computed with an energy spectrum proportional to $E^{-2}$ for the source. BG is background from atmospheric muons. After 
optimizing the analysis parameters, the sensitivity was evaluated for  power law spectra with indices between 2.0 and 3.0. 

\begin{figure} [h]
\centering
\epsfxsize=8cm\epsffile{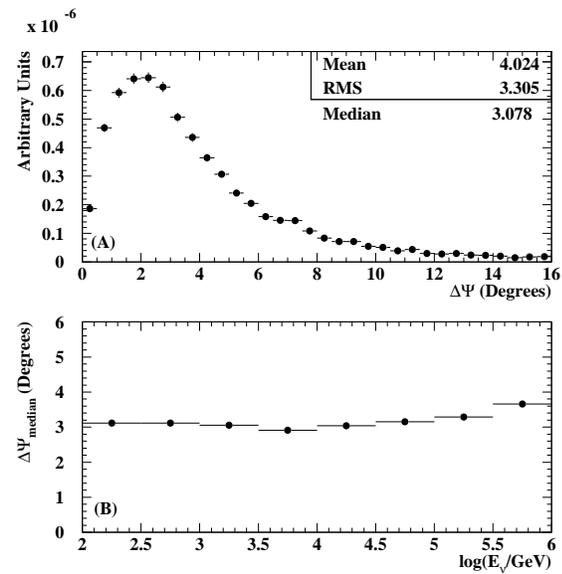}
\caption{Error in the space angle for simulted signal events with energy spectra proportional to ${E_{\nu}}^{-2}$.  Top: distribution of error
averaged over declination; Bottom: Space angle error as a function of neutrino energy.}
\label{fig:angle}
\end{figure}

The space angle resolution is determined from simulation.   The upper panel of
Figure~\ref{fig:angle} shows that the median resolution is 3 degrees, and the lower panel indicates that this value only weakly depends on energy. Two studies were used to check the angular resolution and absolute offset.  First, events that simultaneously trigger the GASP ACT\cite{GASP} and AMANDA provide a "test beam" containing single muons with directional information provided by GASP.   To improve the statistical accuracy of the investigation, a second study involved events which simultaneously trigger the SPASE air shower array\cite{SPASE} and AMANDA.  Although the interpretation of these special events is complicated by the presence of multiple muons, which tend to reconstruct with worse angular precision than single muon events, the response of the detector to these events appears to be correctly modeled.  

The point source analysis yields an event sample of 1097 events
which are distributed on the sky as shown (Fig.\ref{fig:sky}):
\begin{figure} [h]
\centering
\epsfxsize=8cm\epsffile{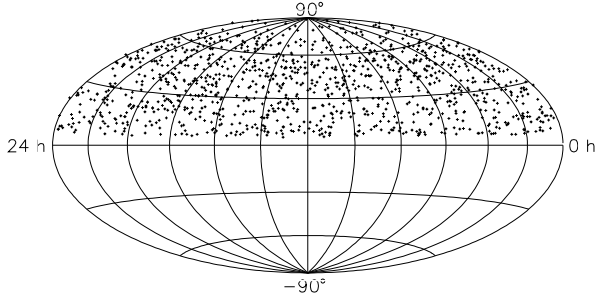}
\caption{Sky distribution of 1097 events in point source analysis. Coordinates are Right Ascension (RA) and declination (dec).}
\label{fig:sky}
\end{figure}

Guided by the estimate of angular resolution, the sky was divided into 319 non-overlapping angular bins. 
The distribution of counts per sky bin is 
consistent with random fluctuations, which were determined by 
selecting all events within a declination band and randomly redistributing them in Right Ascension.  

\begin{figure} [h]
\centering
\epsfxsize=8cm\epsffile{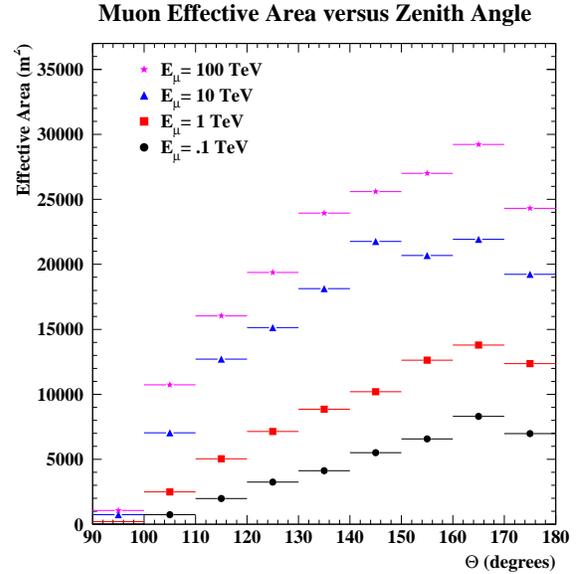}
\caption{ The effective area for muon detection as a function of zenith
angle for $E_{\mu}$ between 0.1 TeV and 100 TeV ($180^{o}$is vertically up in local detector coordinates).}
\label{fig:area}
\end{figure}

\begin{figure} [h]
\centering
\epsfxsize=8cm\epsffile{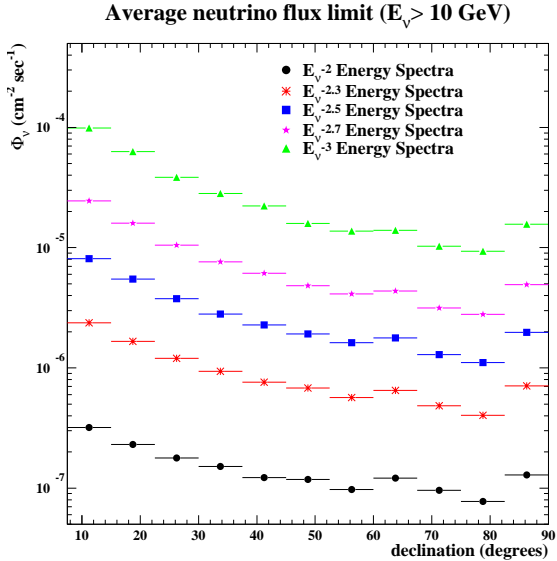}
\caption{ Preliminary neutrino flux limit (90\% CL) on point sources of high energy neutrinos as a 
function of declination, averaged over RA. The limit is computed for a lower energy threshold is 10 GeV. Note that the power law exponent refers to the neutrino energy spectrum.  Also, neutrino absorption by the earth is taken into account.}
\label{fig:limit}
\end{figure}

The neutrino limits were computed according to
\begin{equation}
\phi_{\nu}^{limit} (E_{\nu} > E_{\nu}^{min} ) =
\frac{\mu ( N_{b}, N_{0} ) }{T_{live} \cdot \epsilon \cdot 
\overline{A}_{eff}^{\nu}} \label{eq:fluxlimit2}
\end{equation}

where $\overline{A}_{eff}^{\nu}$, is the neutrino effective area weighted by the assumed neutrino energy spectrum. This quantity is related to the
muon effective area shown in Figure~\ref{fig:area}. The factor $T_{live}$ is
the livetime, and $\epsilon$ is the efficiency due to finite angular 
resolution and also accounts for non-central source placement within an
angular bin. The term
$\mu ( N_{b}, N_{0} )$ generates the 90\% CL according to Feldman and Cousins\cite{stat} for signal events given the measured number of events in the bin, $N_{0}$ and the expected background $N_{b}$ determined from the events in the declination band containing the source bin. The results of this calculation are shown in Figure~\ref{fig:limit} for various assumed spectral indices.

The inferred limits on neutrino flux apply to point sources
with continuous emission (or episodic emission averaged 
over a time interval of 
approximately 0.6 years) and power law energy spectra with a fixed
spectral index above the energy threshold of the detector.  The
limits for sources at large positive declination are comparable to the best published limits in the Southern sky\cite{MACRO99}.

\section{ATMOSPHERIC NEUTRINOS}
\begin{figure} [h]
\centering
\epsfxsize=8cm\epsffile{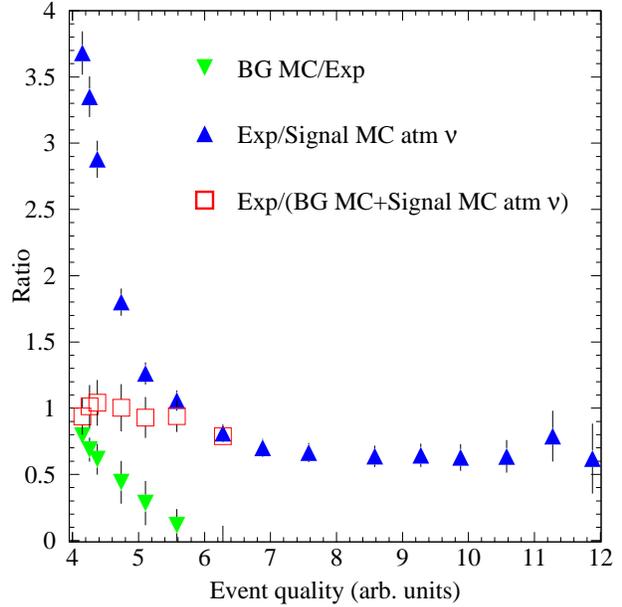}
\caption{ Ratios of passing rates for simulated background (BG MC),
simulated atmospheric neutrino signal (Signal MC atm. $\nu$), and reconstructed experimental data (Exp) as a function of "event quality", a variable which measures the severity of the selection criteria.}
\label{fig:Atmnu_BG}
\end{figure}

The final event sample in the search for point sources contains both atmospheric neutrinos and poorly reconstructed downgoing muons.  The fraction of atmospheric $\nu$ in the sample can be enhanced at the expense of sensitivity. Experimental data is dominated initially by background events -
typically downward going atmospheric muons with poorly known directions.  This can be seen in figure~\ref{fig:Atmnu_BG}.
as indicated by the flat behavior for less restrictive selection criteria
(quality $<4.2$). As selection criteria become progressively more restrictive (increasing values along the x-axis), the asymptotic
flattening of the ratio (experimental data)/(Signal MC Atm.$\nu$) indicates that the evolution of the experimental data becomes consistent with signal expectation in the vicinity of the plot where the (BG MC)/Exp ratio approaches zero.  From this evidence (and visual inspection), we conclude that the contamination in the atmospheric neutrino sample from known physics effects is small ($<15$\%) for values of the event quality parameter greater than five. Background simulations with much greater statistical precision are currently underway.

\begin{figure} [h]
\centering
\epsfxsize=8cm\epsffile{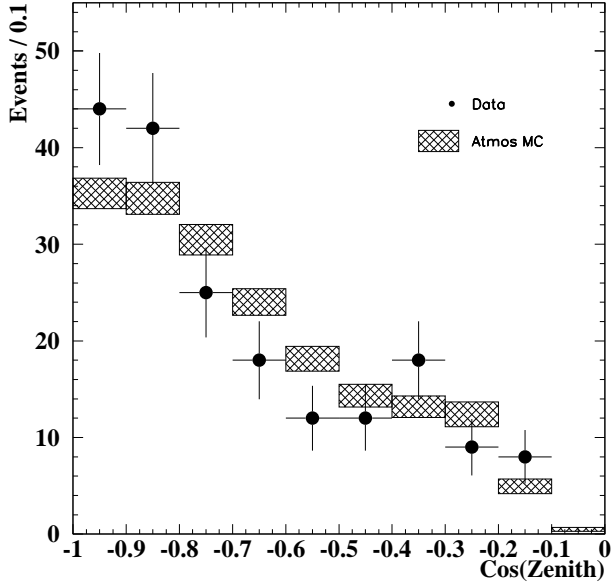}
\caption{Reconstructed zenith angle distribution.  The simulated atmospheric neutrino events (shaded boxes) are normalized to data (filled circles).  The vertical widths of the boxes indicate the errors computed using binomial statistics.}
\label{fig:Atmnu_cos}
\end{figure}

Figure~\ref{fig:Atmnu_cos} shows that the angular distribution of 188 events is also consistent with the simulated distribution of atmospheric $\nu$ events.  Due to the elongated cylindrical geometry of AMANDA-B10, the acceptance shows strong dependence on zenith angle.  

Thus, the angular dependence of the atmospheric neutrino sample is consistent with expectation and contamination from background is small. Finally, the distribution of the number of OMs in an event is also consistent with expectation (see figure~\ref{fig:diffuse}). 

\section{WIMP SEARCH}
\begin{figure} [h]
\centering
\epsfxsize=8cm\epsffile{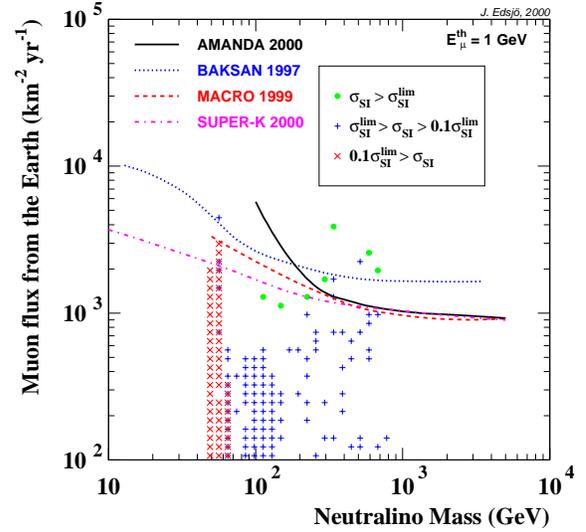}
\caption{High energy muon flux predictions and experimental limits due to the annihilation of supersymmetric particles in the center of the earth. See ref. \cite{Edsjo} for explanation of symbols. The AMANDA limit (90\% C.L.) was corrected to correspond to a threshold of 1 GeV. }
\label{fig:WIMP}
\end{figure}

There is strong evidence for a non-baryonic component in the dark matter in the Universe. Supersymmetric extensions to the Standard Model provide promising dark matter candidates, such as the neutralino. Dark matter in the galactic halo can interact with nuclei in the earth, lose energy and become trapped. The annihilation of neutralinos within the core of the earth will generate high energy neutrinos in the nearly vertical direction (cos$(\theta)<-0.97$). For this highly restricted search cone, the number of events observed agrees with prediction (17 predicted, 15 observed).  Therefore, the lack of a statistically significant excess of neutrino events in the nearly vertical direction can be used to constrain the flux of neutrinos from neutralino annihilation. Figure~\ref{fig:WIMP} compares the AMANDA limits with predictions\cite{Edsjo} for a broad class of supersymmetric models, and with curves derived\cite{Joakim} from published limits by MACRO\cite{MACRO99} and Baksan\cite{Baksan},  illustrating the potential of the technique.
Systematic errors associated with detector sensitivity and neutrino oscillation physics are currently under investigation, but are expected to weaken the limits. 

\section{HIGH ENERGY DIFFUSE FLUX}

\begin{figure} [h]
\centering
\epsfxsize=8cm\epsffile{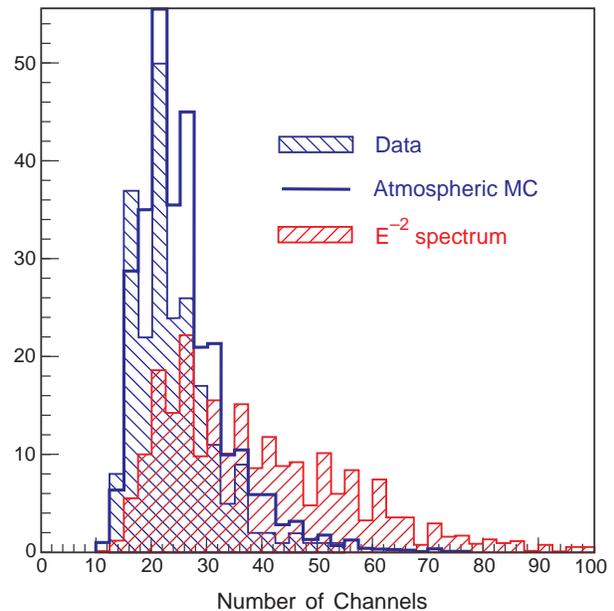}
\caption{Distribution of the number of optical modules for the data presented in Fig.~\ref{fig:Atmnu_cos}. The MC simulation for atmospheric and a generic spectrum $10^{-5}E^{-2} cm^{-2}s^{-1}sr^{-1}GeV^{-1}$ are also shown. }
\label{fig:diffuse}
\end{figure}

Many models of diffuse emission from unresolved sources predict an energy spectrum that is much harder than the atmospheric neutrino spectrum.  Therefore, a limit on the diffuse flux can be extracted by searching for an excess of events at large energies. A simple, although not very precise, measure of energy of the neutrino-induced muons is counting the number of hit channels - optical modules that detected cherenkov light. Figure~\ref{fig:diffuse} shows the distribution of the hit multiplicity for experimental data, for simulated atmospheric neutrinos, and for an arbitrarily normalized, relatively hard energy spectrum. The experimental data agree with the atmospheric neutrino spectrum.  From the non-observation of an excess of high multiplicity events, we derive an upper limit on an assumed diffuse $E^{-2}$ spectrum.  The preliminary limit (90\% CL) is of order $dN/dE_{\nu}\leq 10^{-6}{E_{\nu}}^{-2}cm^{-2}s^{-1}sr^{-1}GeV^{-1}$.  This experimental limit is below several models (e.g., ref. \cite{SS96}), although still above the recent bound established by Waxman and Bahcall\cite{WB99}. The experimental limit can be improved by developing better energy estimators using charge and topological distributions. The response of the detector to muons with energies in excess of 100 TeV is still under investigation.

\section{GRB SEARCH}
Gamma Ray Bursts (GRBs) are thought to be produced by highly relativistic outflows originating from a compact, explosive event. A search for emission of high energy neutrinos correlated in time and direction with GRBs was performed.  We examined a total of 78 GRBs detected by BATSE. The $T_{90}$ duration of the bursts, defined as the time required to accumulate between 0.05-0.95 of the total counts,  varied between 0.2s and 200s.  Guided by several models for neutrino emission\cite{WB}\cite{HJ}, the analysis procedure optimized the effective area of the AMANDA detector for neutrino energies between 100 TeV and 1 PeV.   The lack of a statistically significant excess from any GRB examined in 1997 leads to a range of limits (the variation is primarily due to attenuation by the earth which depends on arrival direction) on the fluence that span from ${E_{\nu}}^2dN/dE_{\nu}\leq 0.05-1.0\times10^{-1}\times  min(1,E_{\nu}/E_{break})TeV cm^{-2} $.  The parameter $E_{break}$, which is of order 500 TeV, is the energy that characterizes the change in the power law of neutrino spectrum (approximately $E^{-1}$ for $E_{\nu}<E_{break}$ and $E^{-2}$ for neutrinos with energies larger than $E_{break}$. Another change in the spectral index is expected at even higher energies, but this feature does not impact our analysis. Using all of the GRBs, a preliminary cumulative upper limit of ${E_{\nu}}^2dN/dE_{\nu}\leq 4\times 10^{-4}min(1,E_{\nu}/E_{break})TeV cm^{-2}$ was also derived\cite{Bay00}. For comparison, Waxman and Bahcall\cite{WB} predict a fluence of $4.8\times 10^{-7}min(1,E_{\nu}/E_{break})TeV cm^{-2}$.


\section{ACKNOWLEDGEMENTS}
This research was supported by
the U.S. National Science Foundation Office of Polar Programs
and Physics Division,
the University of Wisconsin Alumni Research Foundation,
the U.S. Department of Energy,
the Swedish Natural Science Research Council,
the Swedish Polar Research Secretariat,
the Knut and Alice Wallenberg Foundation, Sweden,
the German Ministry for Education and Research, the U.S. National Energy Research Scientific Computing Center (supported by the Office of Energy Research of the U.S. Department of Energy), UC-Irvine AENEAS Supercomputer Facility,  Deutsche Forschungsgemeinschaft (DFG). C.P.H. received support from the EU 4th framework of Training and Mobility of Researchers, contract ERBFMBICT91551 and D.F.C. acknowledge the support of the NSF CAREER program.  P.L. was supported by grant from the Swedish STINT program.  P.D was supported by the Koerber Foundation.


\begin{thebibliography}{99}

\bibitem{ICRC99} See contributions in Proc. 26th Inter. Cosmic Ray
Conf.(ICRC99), Salt Lake City, UT, (Aug 1999). HE 3.1.06, HE 6.3.07, 
HE 4.2.06, HE 6.3.01, HE 4.1.15, HE 5.3.05, HE 4.2.05, HE 6.3.02, HE 4.1.14,
HE 4.1.14, HE 3.2.11, HE 4.2.07, HE 5.3.06 at 
krusty.physics.utah.edu/~icrc1999/proceedings.html

\bibitem{point_pred} F. W. Stecker and M.H. Salamon, Space Sci.Rev. 75 (1996) 341-355; J. P. Rachen and P. Meszaros, Phys. Rev. D58 (1998)123005;  V.J. Stenger, J.G. Learned, S. Pakvasa, and X. Tata, ed., Proc.High Energy Neutrino Astrophysics Workshop(U. Hawaii), World Scientific, Singapore; Szabo and Protheroe.


\bibitem{WB99} E. Waxman and J. N. Bahcall, Phys. Rev. D59 (1999) 023002.

\bibitem{Aman99} R. Wischnewski,{\it et al.}, Nucl. Phys. Proc. Suppl. 75A, 412-414, 1999; 

\bibitem{Andres00} E. Andres {\it et al.}, Astropart. Phys., 13(2000)1-20.

\bibitem{TAUP99} S. W. Barwick, for the AMANDA collaboration, Nucl. Phys. B(Proc. Suppl.)87 (2000) 402. 


\bibitem{Bar99} S. W. Barwick, Physica Scripta. T85 (2000),106 (astro-ph/9903467).

\bibitem{MACRO99} T. Montaruli,{\it et al.}, Proc. 26th ICRC (1999), HE 4.2.03.; M. Ambrosio, \textit{et al.}, submitted to Astrophys. J. , (2000) astro-ph/0002492

\bibitem{Mannheim} K. Mannheim, R. J. Protheroe, J.P. Rachen, Phys.Rev D (1999), astro-ph/9812398.

\bibitem{atm_nu_pred}V. Agrawal, T. Gaisser, P. Lipari, and T. Stanev, Phys. Rev. D53(1996)1314.

\bibitem{Nellen} L. Nellen, K. Mannheim, P.L. Biermann, Phys. Rev. D47(1993) 5270.

\bibitem{SS96} F. W. Stecker, M. H. Salamon, Sp. Sci. Rev. 75(1996) 341.

\bibitem{Mannheim93} K. Mannheim, Phys. Rev D48(1993)2408

\bibitem{Cola} S. Colafrancesco, and P. Blasi, Astropart. Phys. 9(1998) 227.

\bibitem{Bednarek97} W. Bednarek and R.J. Protheroe, Mon. Not. Roy. Astro. Soc. 287(1997) 560.

\bibitem{Gaisser98} T.K. Gaisser, R.J. Protheroe, and T. Stanev, Ap. J. 492(1998) 219.

\bibitem{WB} E. Waxman and J.Bahcall, Phys. Rev. Lett. 78(1997)2292.

\bibitem{HJ} F. Halzen and G. Jaczko, Phys. Rev. D54 (1996)2774.

\bibitem{Bay00} R. Bay, Ph.D dissertation, University of California-Berkeley, 2000[astro-ph/0008255].

\bibitem{ML00} J.G. Learned and K. Mannheim, to appear in Ann. Rev. Nucl. Part. Phys.(2000).

\bibitem{Young00}  S. Young, Ph.D. dissertation, University of California-Irvine (2000) to be published.

\bibitem{Edsjo} L. Bergstrom, J. Edsjo, and P. Gondolo,  Phys. Rev. D58(1998)103519 [hep-ph/9806293].

\bibitem{Joakim} J. Edsjo, private communication, 2000.

\bibitem{Baksan} M. M. Boliev, et al., Nucl. Phys. B(Proc. Suppl.)48(1996)83.

\bibitem{stat} G. J. Feldman and R. D. Cousins, Phys. Rev. D57(1998)3873.

\bibitem{SPASE} J. E. Dickinson, et al., Nucl. Inst. Meth. A 440(2000)95.

\bibitem{GASP} G. Barbagli, et al., Nucl. Phys. B(Proc. Suppl.)32 (1993)156. 

\end{thebibliography}
\end{document}